\documentclass[twocolumn,journal]{IEEEtran}
\usepackage[dvips]{graphicx}
\usepackage[english]{babel}
\usepackage{algorithm}
\usepackage{algorithmic}
\usepackage{amsfonts}
\usepackage{booktabs}
\usepackage{multirow}
\usepackage{stfloats}
\usepackage[numbers,sort&compress]{natbib}
\usepackage{amsmath}
\usepackage{amssymb}
\usepackage{array}
\usepackage{url}
\usepackage{color}

\usepackage{epsfig,latexsym}
\usepackage{flushend}

\title{\huge Enabling 5G on the Ocean: A Hybrid Satellite-UAV-Terrestrial Network Solution}

\author{
\IEEEauthorblockN{Xiangling~Li, Wei~Feng, Jue~Wang, Yunfei~Chen, Ning~Ge, Cheng-Xiang~Wang}\\
\thanks{X.~Li, W.~Feng (corresponding author), and N. Ge are with Tsinghua University; J.~Wang is with Nantong University; Y. Chen is with University of Warwick; C.-X.~Wang is with Southeast University and Purple Mountain Laboratories. W.~Feng and J.~Wang are also with the Peng Cheng Laboratory.}
}

\begin{document}
\maketitle

\begin{abstract}
Current fifth generation (5G) cellular networks mainly focus on the terrestrial scenario. Due to the difficulty of deploying communications infrastructure on the ocean, the performance of existing maritime communication networks (MCNs) is far behind 5G. This problem can be solved by
using unmanned aerial vehicles (UAVs) as agile aerial platforms to enable on-demand maritime coverage, as a supplement to marine satellites and shore-based terrestrial based stations (TBSs). In this paper, we study the integration of UAVs with existing MCNs, and investigate the potential gains of hybrid satellite-UAV-terrestrial networks for maritime coverage. Unlike the terrestrial scenario, vessels on the ocean keep to sea lanes and are sparsely distributed. This provides new opportunities to ease the scheduling of UAVs. Also, new challenges arise due to the more complicated maritime prorogation environment, as well as the mutual interference between UAVs and existing satellites/TBSs. We discuss these issues and show possible solutions considering practical constraints.
\end{abstract}

\IEEEpeerreviewmaketitle

\section{Introduction}
With the continuous development of marine activities, the demand for maritime broadband communications increases dramatically. Till now, the data rate that can be supported on the ocean has approached a few Mbps \cite{wei_survey}. However, this is still way below that supported by the fifth generation (5G) cellular network at the scale of Gbps. To meet the dramatically increasing demand, new solutions to maritime communication networks (MCNs) have become a pressing need.

Different from the urban area, it is challenging to densely deploy base stations on the ocean. In order to extend 5G services to the ocean, Ericsson and China Mobile jointly established a Time Division Long Term Evolution (TD-LTE) trial network in the Qingdao sea area of China. By building
shore-based terrestrial based stations (TBSs) along the coast, this trial network can provide broadband communication services for an area of up to tens of kilometers away from the shore \cite{TBS}. To further extend the coverage, multi-hop system was adopted in the TRITON \cite{TRITON} and BlueCom+ projects \cite{BLUECOM}. In these projects, vessels were employed as relay nodes to enhance communications. Moreover, tethered balloons at an altitude of 120 m were utilized in the BlueCom+ project to further enhance the coverage of vessel-based relays. It can offer data rates
in excess of 3 Mbps up to 150 km offshore. However, as most vessels follow fixed sea lanes to avoid shipwrecks, this multi-hop solution lacks flexibility. Coverage holes may exist in areas far away from the sea lanes of the relay nodes.

To cover more remote areas far away from the coast, satellites can be exploited. The most well-known solution is the marine satellite, i.e., the Inmarsat. Because of its inherent long transmission distance, the data rate of satellite communications is usually much less than that of the terrestrial 5G. In order to meet the increasing data demand, developing high-throughput satellites has attracted much research attention.
For example, the Inmarsat's fifth-generation (Inmarsat-5) satellite network deployed in the Geostationary Earth Orbit (GEO) can offer Ka-band services of 50 Mbps forward and several Mbps return data rates \cite{Inmarsat_5}. Besides, the Iridium NEXT system consisting of 66 Low Earth Orbit (LEO) satellites at an altitude of 780 km is expected to offer Ka-band services with data rates of up to 8 Mbps \cite{Iridium_NEXT}.
These efforts have substantially improved the performance of satellite communications. However, the large communication delay remains an open issue. Moreover, these new developments require dedicated terminals using high gain antennas.

\begin{figure*}[!t]
  \centering
  \includegraphics[width=6 in]{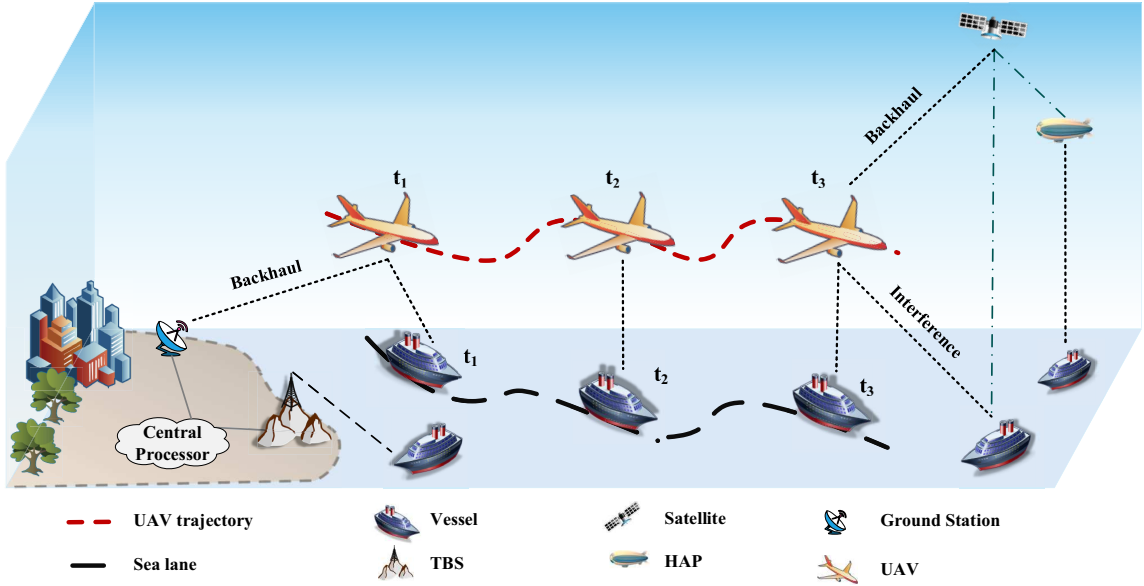}%
  \caption{Illustration of a hybrid satellite-UAV-terrestrial network for agile broadband maritime coverage.}\label{fig_simulation_system}
\end{figure*}

In addition to shore-based TBSs and marine satellites, high-altitude aerial platforms (HAPs) can also be used as communications infrastructure. For example, the Loon project employs super-pressure balloons at an altitude of around 20 km to realize broadband coverage for countryside and remote areas \cite{Loon}. This network was reported to provide communication services of up to 10 Mbps.
Also, as aerial communication platforms, unmanned aerial vehicles (UAVs) are more agile than balloons, due to their better mobility at a lower altitude. Most existing studies on UAV communications focus on the terrestrial scenario, where it has been
recognized that
UAVs are promising for dynamic coverage enhancement \cite{zeng_magazine,feng_iot}.
Although the maritime environment is quite different from terrestrial scenarios, one believes that UAVs can also offer agile
aerial platforms above the ocean to enable on-demand maritime coverage enhancement.

In this article, we investigate the integration issue of UAVs with existing MCNs. In the concerned scenarios, UAVs are flexibly deployed to fill up the broadband coverage holes on the ocean, which cannot be covered by conventional shore-based TBSs and marine satellites.
The integration leads to a hybrid satellite-UAV-terrestrial network architecture. We investigate the potential gains of hybrid satellite-UAV-terrestrial networks for maritime coverage in the 5G era. Different from existing studies on UAV communications, vessels are the main users on the ocean, which often keep to sea lanes for safety and are sparsely distributed. These properties render it possible to
elaborately schedule the UAVs to match the user demand. Within this framework, we also discuss new challenges of deploying UAVs above the ocean, which stem from the more complicated maritime prorogation environment, as well as the mutual interference between UAVs and existing shore-based TBSs/satellites.

\section{Opportunities for Maritime UAV Communications}
\subsection{Agile Mobility of UAVs}
As illustrated in Fig.~\ref{fig_simulation_system}, a hybrid satellite-UAV-terrestrial maritime network can be established by integrating UAV communications into existing MCNs.
In contrast to on-shore TBSs and satellites, the unique advantage of UAVs lies in their agile mobility.
In the following, we compare the mobility of TBS, satellites and UAVs.

\begin{itemize}
  \item \textbf{TBS}. In general, TBS should be deployed on the mountains or highly-elevated towers along the coastline. Thus, the deployment of TBSs is quite limited and fixed in practice. To enhance the mobility, shipborne base stations may be deployed, which play a similar role as the TBSs. However, their mobility remain limited due to the restriction of sea lanes. This restriction cannot be broken in general, because it concerns the navigation safety of the corresponding vessel.
    \item \textbf{Satellite}. According to the Orbital Dynamics theory, the deployment of satellites is largely restricted. For instance,
     both the aforementioned Inmarsat-5 and Iridium NEXT satellites follow certain orbits mainly determined by astrodynamics.
     In general, we are able to choose a proper orbit, but cannot create an arbitrary orbit.
     For this reason, the expensive LEO constellation is usually necessary to achieve global coverage.
  \item \textbf{UAV}. The UAV has the most flexible deployment. As shown in Fig.~\ref{fig_simulation_system}, a UAV can fly with the target vessel, so as to provide on-demand broadband communication services. Nevertheless, the endurance of UAVs is usually limited because of limited energy onboard. Likewise, the weather condition also imposes restrictions on the deployment of UAVs. Therefore, it is necessary
      to optimize the scheduling of UAVs considering all practical constraints.
\end{itemize}

In summary, the agile mobility of UAVs is unique and quite valuable, because the access point equipped at UAV can fly closer to the target user, thereby significantly improving the transmission rate and shortening the communication latency. By exploiting the characteristics of maritime
user distribution and predictable mobility as described in the following, it is possible to improve the UAV efficiency in maritime communications.

\subsection{Unique Characteristics of Maritime Users}
\begin{figure}[!t]
  \centering
  \includegraphics[width=3.5in]{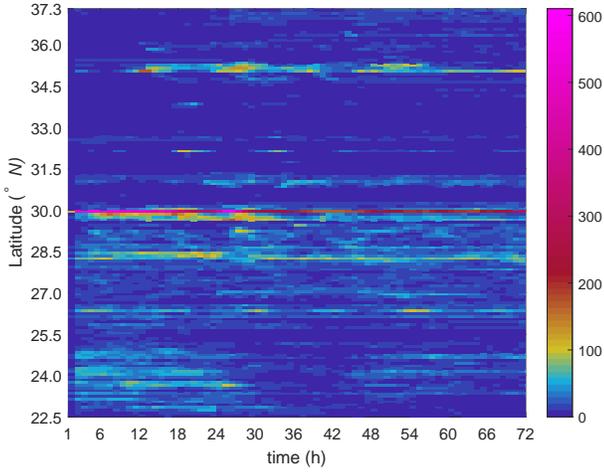}%
  \caption{Illustration of the practical vessel distribution within a coastal area of China, where the offshore distance is in the range of [20,~30] km, and the time duration is from 1st October 2015 to 3rd October 2015.}\label{fig_distri_ships_over_time}
\end{figure}

Different from the terrestrial case where the majority of users move randomly, vessels on the ocean have unique characteristics in terms of both distribution and mobility.
Their distributions are both spatially and temporally sparse on the vast ocean.
As an example, we show the typical vessel distribution within a coastal area of China in Fig. \ref{fig_distri_ships_over_time}.
The practical Automatic Identification System (AIS)~\footnote{AIS is a transponder system for ships intending to increase their safety. An AIS transmitter regularly reports the ship's state information, e.g., the position, heading, speed and so on.} data is used to obtain the distribution.
For the spatial domain, the latitude is in the range of $[22.5^\circ N,~ 37.3^\circ N]$ and the offshore distance is in the range of [20,~30] km.
For the temporal domain, a period from 1st October 2015 to 3rd October 2015 is taken into account.
In the figure, the number of vessels appeared in a square area with latitude $0.1^\circ$ in length and 10 km in width during an hour is accumulated as one data point. It is shown that vessels are sparsely distributed in both spatial and time domains.
For most of the areas, the color map is dominated by dark blue, indicating that very few users (or even no user) are distributed in these areas. The red line around latitude $30^\circ$ indicates the existence of a sea lane.

\begin{figure}[!t]
  \centering
  \includegraphics[width=3.5in]{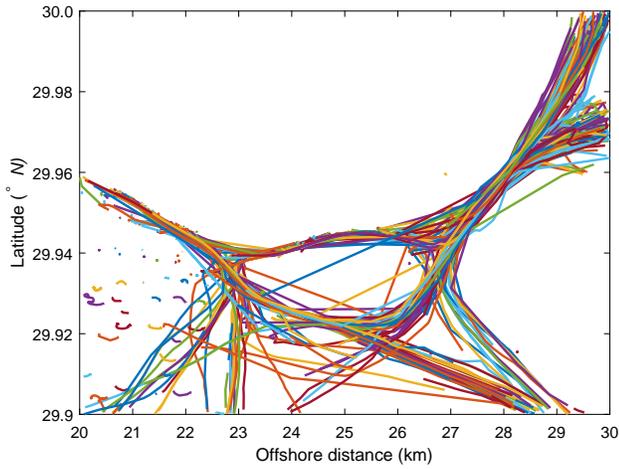}%
  \caption{Illustration of sea lanes within a coastal area of China, cumulated from the practical AIS data of 610 vessels during one hour.}
  \label{lane}
\end{figure}

Actually, most maritime users follow fixed shipping lanes rather than randomly move. We further illustrate various sea lanes
in Fig. \ref{lane}. The curves in the figure are obtained from 610 vessels during one hour using the same AIS data as Fig. \ref{fig_distri_ships_over_time}. The latitude is in the range of $[29.9^\circ N,~ 30.0^\circ N]$ and the offshore distance is in the range of [20,~30] km. Although there exist some randomly moving vessels as shown in the left-bottom area of the figure, the majority of vessels keep to sea lanes, and have regular and predictable mobility patterns.
Thus, these users can be easily tracked despite of the vastness of ocean area. This creates opportunities for the efficient scheduling of UAVs.

\begin{table*}[!t]
\renewcommand{\arraystretch}{1.2}
\small
  \centering
  \caption{A summary on the key parameters of typical UAV products.}
  \label{table_2}
\begin{tabular}{|l|l|l|l|l|l|l|l|l|}
  \hline
  \multicolumn{1}{|m{0.40\columnwidth}|}{\textbf{Type}}            & \multicolumn{3}{c|}{\textbf{\emph{Unmanned gyroplane}}} & \multicolumn{3}{m{0.55\columnwidth}|}{\centering \textbf{\emph{Vertical take-off and landing fixed-wing UAV}}}  & \multicolumn{2}{c|}{\textbf{\emph{Helicopter}}}  \\
  \hline
  \multicolumn{1}{|m{0.40\columnwidth}|}{\textbf{Company }}                 & DJI            & CEEWA         & Rising fly    & JOUAV     &  ZEROTECH                 & JOUAV            & Ziyan UAV     & AEE \\
  \hline
  \multicolumn{1}{|m{0.40\columnwidth}|}{\textbf{Model}}                    & Inspire 2      & X-9           & MixOne        & CW10            &  ZT-30V                   & CW100     & Blowfish A2   & AU300\\
  \hline
  \multicolumn{1}{|m{0.40\columnwidth}|}{\textbf{Wind resistance (m/s)}}    & 10             & 12            & /             & 10.8-13.8   &  10.8-13.8                & 13.9-17.1            & 17            & /  \\
  \hline
  \multicolumn{1}{|m{0.40\columnwidth}|}{\textbf{Cruising speed (km/h)}}  & /              & /             & /             & 72               &  90                       & 100      & 70-90       & 130  \\
  \hline
  \multicolumn{1}{|m{0.40\columnwidth}|}{\textbf{Maximum speed (km/h)}}  & 94    & 72         & /         & / &  130               & /                 & 130           & /   \\
  \hline
  \multicolumn{1}{|m{0.40\columnwidth}|}{\textbf{Maximum duration of flight (h)}}   & 0.5            & 1             & 5          & 1.5       &  7                        & $4-8$     & 1          & 4   \\
  \hline
  \multicolumn{1}{|m{0.40\columnwidth}|}{\textbf{Driving force}}          & electric    & electric  &
  \multicolumn{1}{m{0.17\columnwidth}|}{oil-electric hybrid}   & electric    &
  \multicolumn{1}{m{0.18\columnwidth}|}{oil-electric hybrid}   & \multicolumn{1}{m{0.12\columnwidth}|}{oil}         & electric   & \multicolumn{1}{m{0.1\columnwidth}|}{oil}  \\
  \hline
\end{tabular}
\end{table*}

\subsection{On-Demand Coverage by Maritime UAVs}
Exploiting both the agile mobility of UAVs and the unique characteristics of vessels, it is natural to design an on-demand coverage framework.
For a vessel user that requires broadband services outside the coverage area of existing MCNs, a UAV can be dispatched.
The UAV can either work in a serve-and-leave manner, or it can move with the vessel to guarantee long-term broadband services.
After the transmission task has been accomplished, the UAV flies back to the charging station, or towards the next vessel user in the service queue.
This is quite different from conventional UAV communications where users are assumed to be fixed or have random distribution and moving patterns.

Compared with TBSs and satellites, which can also support on-demand coverage with dynamic beams at the cost of expensive antenna arrays,
the mobile agility of UAVs makes it possible to accomplish this in a more efficient way.
In particular, UAVs can be dynamically deployed only to cover the sea lanes.
In the temporal domain, communication requests could appear intermittently. Then, UAVs can be flexibly and dynamically scheduled according to the time-varying communication demand. These imply that UAV-enabled maritime on-demand coverage has great potential to improve the efficiency of maritime communications.

\section{Challenges and Possible Solutions}
In this section, we discuss the challenges of integrating UAVs into existing MCNs. They are summarized by the following three aspects: 1) harsh maritime environment may affect the real-time deployment of UAVs, 2) the hybrid network architecture requires joint resource allocation and interference coordination, and 3) limited channel state information due to the dynamic propagation environment and large transmission delay will bring new challenges to the system optimization.

\subsection{Harsh Maritime Environment}
Different from the terrestrial case, the maritime environment is seriously affected by weather conditions, such as typhoon and disastrous waves. In the extreme case, the wind speed caused by typhoon could be larger than 30 m/s. Most existing UAV products are not
designed for all-weather service. As summarized in Table \ref{table_2}, they are more likely to be deployed under relatively good weather conditions, i.e., wind speed smaller than $17.1$ m/s. In practice, the UAV used in the maritime environment should be carefully chosen.

Another important issue is that the vast sea area makes it difficult for UAVs to land and charge, which seriously restricts the UAV deployment in practice. Offline deployment of UAVs taking these restrictions into account is a possible solution. As discussed in the previous section, most vessels travel regularly along sea lanes. This can provide important prior information for the deployment of UAVs. For example, by using the historical information on the communication demand over sea lanes, the hotspot areas where broadband coverage is requested can be predicted. By intentionally deploying UAVs within their endurance time over these areas, broadband coverage holes can be efficiently filled. Also, the serving latency can be reduced by this pre-deployment scheme in contrast to the request-triggered temporary dispatch manner. In practice, the time advance and duration of offline deployments should be controlled within the predictable range of maritime environment. Online decision should also be activated for better adaptability in extremely dynamic weather conditions. This leads to an online and offline collaboration framework.

As summarized in Table \ref{table_2}, the UAV's maximum duration of flight is usually less than 8 hours due to the limited energy onboard. The UAV deployment should be carefully determined according to the residual energy, and how to deploy service stations for energy replenishment on the vast ocean area becomes an important issue. To address this problem, vessels can be used as service stations. But as discussed above, their locations are restricted to sea lanes. Thus, dedicated and vessel-based service stations should be deployed in a synergetic manner. Note that the offshore distance of the coastal area is about 370 km for the exclusive economic zone. If service stations are only deployed along the coast, considering the cruising speed and maximum flight time, only the oil-powered fixed-wing UAV is possible for a 740 km round trip from Table \ref{table_2}. The other UAVs listed in Table \ref{table_2} can only work in areas near the coast if vessel-based service stations are not available.

To achieve continuous communication using the energy-limited UAVs, efficient scheduling of a warm of UAVs is necessary. Recalling that vessels are sparsely distributed, it is more likely that a maritime UAV could be part-time idle during its flying time. Hence, when a UAV has to go to terrestrial/shipborne service stations for energy replenishment, neighbouring idle UAVs with enough residual energy can be dispatched as replacements to guarantee the continuous coverage. This leads to another important optimization dimension of UAV scheduling, that is minimizing the number of UAVs scheduled, which not only saves costs but also facilitates management. In the extreme case that there are no neighbouring idle UAVs with enough residual energy, vessels may
temporarily request degraded services from existing MCNs.

\subsection{Coordination Issues}
Maritime UAVs are part of a hybrid satellite-UAV-terrestrial communication network. They rely on existing MCNs for backhaul links.
Different from traditional UAV communications in the cellular architecture, where the backhaul is not crucial due to the ubiquitous coverage of cellular networks, current MCN is usually not sufficient to build a reliable wireless backhaul for UAVs on the vast ocean. Specifically, TBSs can only support UAVs in the coastal area. When UAVs are far away from the coast, satellites could be the only choice for wireless backhaul with
inevitable large delay and limited communication rate. Moreover, to communicate with satellites, UAVs should be equipped with airborne high-gain antennas. Considering these facts, the backhaul issue should be taken into account in the scheduling of UAVs. Alternatively, data caching on the UAV can be used, which allows interim outage of backhaul given the information delay tolerance. In this case, communications, control of UAV's trajectory and caching need to be jointly designed.

In addition to backhaul, UAVs may also share spectrum with existing MCNs so as to alleviate the spectrum scarcity problem.
However, due to the mobility of UAVs, the co-channel interference under spectrum sharing is more complicated than the traditional case with fixed communications infrastructure \cite{Cognitive}. In practice, the trajectory of UAVs can be exploited to predictively characterize the interference
distribution. By doing so, process-oriented interference coordination can be derived between UAVs and TBSs/satellites.

\subsection{Limited Channel State Information}
To improve the quality of service, the location (or trajectory) planning and the resource allocation for UAVs are required using the channel state information (CSI). However, in the maritime scenario, the CSI is usually difficult to acquire due to the following reasons.
\begin{enumerate}
  \item As previously discussed, the trajectory planning for maritime UAVs is likely to be pre-determined offline. This means that the trajectory optimization has to be conducted using only the predictable CSI, rather than the instantaneous CSI.
  \item When UAVs share spectrum with satellites (or TBSs) to improve spectrum efficiency, the interference from UAVs to satellite users is inevitable. To mitigate the interference, the CSI between UAVs and satellite users has to be known. However, in practice, there are usually no direct links between UAVs and satellite users for CSI feedback. This CSI has to be exchanged between satellite sub-system and UAV sub-system via a dedicated central processor, which may lead to undesirable delay.
\end{enumerate}

In practice, the large-scale CSI, such as path loss, shadowing, angle of departure, angle of arrival and so on, varies slowly and is closely related to transceiver's positions, which could be predicted by using the historical data and/or pre-measured data \cite{wei_lane}.
To deal with the challenges mentioned above, utilizing large-scale CSI could be a reasonable choice for the optimization of a hybrid satellite-UAV-terrestrial network. We can create a \emph{radio map} on the ocean focusing on shipping lanes. The map generates the large-scale CSI for any given positions. In its initial stage, dedicated UAVs and vessels can be dispatched to measure the large-scale CSI. Then, communication data
containing channel knowledge can be used to update the radio map for better resolution in an online manner.
This creates a novel lookup-table approach for CSI acquisition instead of conventional pilot-based channel estimation and feedback approaches. Correspondingly, new methodology for reliable resource allocation and the placement (or trajectory) optimization for UAVs with large-scale CSI, i.e., a radio map in practice, should be conceived.

\begin{table}[!t]
\renewcommand{\arraystretch}{1.2}
\small
  \centering
  \caption{Simulation Parameters.}
  \label{table_3}
\begin{tabular}{|p{0.5\columnwidth}|p{0.4\columnwidth}|}
  \hline
  \textbf{Parameter}         & \textbf{Value }                  \\ \hline
  Transmit power of UAV     & $[22,~40]$ dBm  \\ \hline
  Transmit power of TBS     & 40 dBm  \\ \hline
  Antenna gain of UAV       & 8 dBi  \\ \hline
  Antenna gain of TBS       & 12 dBi  \\ \hline
  \multicolumn{1}{|m{0.5\columnwidth}|}{Antenna gain of UAV's vessel}          & 8 dBi  \\ \hline
  \multicolumn{1}{|m{0.5\columnwidth}|}{Antenna gain of satelllite's vessel}   & 30 dBi  \\ \hline
  Altitude of TBS           & 100 m  \\ \hline
  Velocity of UAV           & $[20,~36]$ m/s  \\ \hline
  Acceleration of UAV       & $[0,~5]$ $\textrm{m}\textrm{/s}^2$  \\ \hline
  Altitude of UAV           & $[2.6,~5]$ km      \\ \hline
  \multicolumn{1}{|m{0.5\columnwidth}|}{Residual communication energy}    & ${\{1.5\times10^{3},~3\times10^{4}\}}$ J     \\ \hline
  \multicolumn{1}{|m{0.5\columnwidth}|}{Interference temperature limitation} & ${\{-55,~-40\}}$ dBm \\ \hline
  \multicolumn{1}{|m{0.5\columnwidth}|}{Rician \emph{K} factor} & 10 \\
  \hline
  \multicolumn{1}{|m{0.5\columnwidth}|}{Path loss model} & \multicolumn{1}{m{0.4\columnwidth}|}{${L}\left( {{\rm{dB}}} \right) =116.7 + 15\log_{10}\left( {\frac{{{d}}}{{{2.6\times 10^3}}}} \right)$} \\
  \hline
\end{tabular}
\end{table}

\section{Numerical Example and Discussions}
We use an example to show the benefit of hybrid satellite-UAV-terrestrial networking, as illustrated in Fig. \ref{fig_simulation_system}. The UAVs are dispatched in an on-demand manner: a UAV is sent to the objective vessel on request, and flies back to the service station when the transmission is accomplished. The trajectory of the UAV from time $t_1$ to $t_3$ is pre-designed according to the shipping lane information and predicted large-scale channel information. On the backhaul side, the UAV directly communicates with the nearest TBS. On the access side, the UAV shares spectrum with satellites in an opportunistic manner. The interference from the UAV to the satellite users is controlled by an interference temperature limitation $I$. Also, orthogonal resources, e.g., different subcarriers or different time slots, are used to mitigate the interference between the access link and the backhaul link of the UAV.

\begin{figure}[!t]
  \centering
  \includegraphics[width=3.5in]{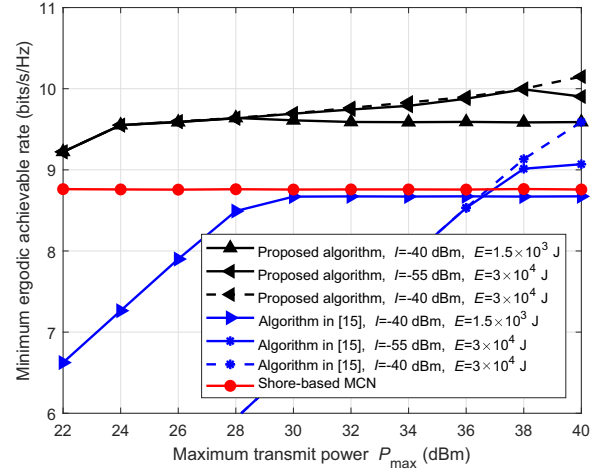}%
  \caption{Minimum ergodic achievable rate of different algorithms.}\label{fig_simulation_result}
\end{figure}

A typical composite channel model is considered, consisting of both path loss and Rician fading \cite{air_ground_channel, channel_survey}. We assume that only the large-scale CSI is available for the UAV pre-deployment. The trajectory and the transmit power of the UAV are jointly optimized to maximize the minimum ergodic achievable rate during the period that the UAV serves the vessel, under various practical constraints including the maximum transmit power ${P_{\max}}$, the residual energy $E$, the limited backhaul capacity, and the interference temperature limitation $I$ \cite{mine_wocc}. The goal of maximizing the minimum achievable rate is to improve the coverage performance, i.e., to promote the worst-case user's performance. Other metrics, e.g., sum rate maximization, can also be pursued according to practical requirements.
We assume that the shipping lane of the vessel is known beforehand. Without loss of generality, we assume the vessel is moving from ${(5.0\times 10^4, 0,10 )}$ m to ${(6.8\times 10^4, 0,10 )}$ m with a velocity of 10 m/s while the UAV serves it. Via simulation, the minimum ergodic achievable rate during the period is compared for different approaches in Fig. \ref{fig_simulation_result}, where the simulation parameter setting is described in Table \ref{table_3}.

First of all, the minimum ergodic achievable rate is compared between the UAV-assisted MCN and the traditional shore-based MCN. For the shore-based MCN, the vessel is directly served by the TBS and we assume that accurate CSI is known at the TBS. Although the UAV-assisted method has additional restrictions, such as backhaul capacity and inaccurate CSI, its performance can still be improved by employing the UAV to
reduce the transmission distance to the vessel.

We also note that it would be inefficient, if not impossible, to directly apply the existing UAV scheduling methods (which was designed for the terrestrial scenario) on the ocean. For comparison, the performances of two UAV scheduling algorithms are demonstrated in Fig.~\ref{fig_simulation_result}, including 1) the algorithm in \cite{zeng_throughput}, which was designed for the terrestrial scenario and has shown significant gains in improving the performance of cellular networks, and 2) the algorithm proposed in \cite{mine_wocc}, which utilizes only the large-scale CSI, and additionally considered constraints on the interference and maximum transmit power. When ${I=-40}$ dBm, the constraint on the interference is looser compared with others, and hence it can be ignored. In Fig. \ref{fig_simulation_result}, when ${P_{\max}\geq 28}$ dBm, ${I=-40}$ dBm and $E=1.5\times 10^3$ J, the performance is not varied when ${P_{\max}}$ is increased, and thus the performance is mainly determined by constraints on the residual energy and the backhaul capacity. Also, when $E=3\times 10^4$ J, the constraint on the residual energy can be ignored. When ${P_{\max}\geq 38}$ dBm and $E=3\times 10^4$ J, the effect of the constraint on the interference can be seen. The performance is improved when ${I}$ is increased. One sees that by using only the large-scale CSI, better performance can be obtained by our tailored algorithm for the maritime applications.

\section{Conclusions}
In this article, we have discussed opportunities and challenges for integrating UAVs into existing MCNs. First of all, we have shown that most vessels keep to sea lanes and are sparse distributed on the ocean. These characteristics of vessels and the UAV's agility bring opportunities to realize the on-demand coverage using UAVs. Moreover, challenges can be well addressed by dynamically deploying and scheduling UAVs, which have been designed considering the coordination among TBSs, UAVs and satellites and the optimization using the predictable large-scale CSI. At last, a case study has been conducted to demonstrate benefits provided by the hybrid satellite-UAV-terrestrial network.

\section*{Acknowledgment}
This work was supported in part by the National Natural Science Foundation
of China (Grant No. 61922049, 61771286, 61941104, 61701457,
61960206006, 91638205); the National Key R\&D Program of China
(Grant No. 2018YFA0701601); the Research Fund of National Mobile Communications
Research Laboratory, Southeast University (Grant No. 2020B01); the EU
H2020 RISE TESTBED2 project (Grant No. 872172); the China Postdoctoral
Science Foundation Project (Grant No. 2019M650680); the Beijing Innovation
Center for Future Chip, and the Peng Cheng Laboratory.

\section*{Biographies}
\noindent
XIANGLING LI received her B.S. and M.S. degrees from Jilin University, Jilin, China, in 2008 and 2011, respectively. She received her Ph.D. degree from Beijing University of Posts and Telecommunications, Beijing, China, in 2017. She is a Postdoctoral Research Fellow at Tsinghua University. Her research interests include maritime broadband communication networks, UAV networks, wireless sensor networks.
\\

\noindent
WEI FENG [S'06, M'10, SM'19] is an associate professor with the Department of Electronic Engineering, Tsinghua University. His research interests include maritime broadband communication networks, large-scale distributed antenna systems, and coordinated satellite-UAV-terrestrial networks. He serves as the Assistant to the Editor-in-Chief of China Communications and an Editor of IEEE Transactions on Cognitive Communications and Networking.
\\

\noindent
JUE~WANG [S'10-M'14] was with the Singapore University of Technology and Design as a Postdoctoral Research Fellow from 2014 to 2016. He is currently with the School of Electronic and Information Engineering, Nantong University, Nantong, China. His research interests include MIMO wireless communications, multiuser transmission, MIMO channel modeling, massive MIMO systems, and physical layer security.
\\

\noindent
YUNFEI CHEN [S'02-M'06-SM'10] received his B.E. and M.E. degrees in electronics engineering from Shanghai Jiaotong University, Shanghai, China, in 1998 and 2001, respectively. He received his Ph.D. degree from the University of Alberta in 2006. He is currently working as an Associate Professor at the University of Warwick, U.K. His research interests include wireless communications, cognitive radios, wireless relaying and energy harvesting.
\\

\noindent
NING GE has been with the Department of Electronics Engineering at Tsinghua University since 2000, where he is a professor and serves as Director of Communication Institute. His research interests include ASIC design, short range wireless communication, and wireless communications. He is a senior member of CIC and CIE.
\\

\noindent
Cheng-Xiang Wang [S'01-M'05-SM'08-F'17] received his Ph.D. degree from Aalborg University, Denmark, in 2004. He has been with Heriot-Watt University, Edinburgh, UK, since 2005, and became a professor in 2011. In 2018, he joined Southeast University, China, as a professor. He is also a part-time professor with the Purple Mountain Laboratories, Nanjing, China. He has co-authored three books, one book chapter, and over 370 papers in refereed journals and conference proceedings. His current research interests include wireless channel measurements/modeling and B5G wireless communication networks. He is a Fellow of the IET, an IEEE Communications Society Distinguished Lecturer for 2019 and 2020, and a Highly Cited Researcher recognized by Clarivate Analytics in 2017-2019.

\begin{thebibliography}{10}
\bibitem{wei_survey}
T. Wei, W. Feng, Y. Chen, C.-X. Wang, N. Ge, and J. Lu, ``Hybrid satellite-terrestrial communication networks for
the maritime internet of things: key technologies, opportunities, and challenges,'' arXiv:1903.11814, Mar. 2019.
\bibitem{TBS}
Ericsson: Maritime ICT Cloud enables ships to join the Networked Society, 2015. [Online]. Available: https://www.ericsson.com/en/press-releases/2015/1/maritime-ict-cloud-enables-ships-to-join-the-networked-society.
\bibitem{TRITON}
M. Zhou, V. D. Hoang, H. Harada, J. S. Pathmasuntharam, H. Wang, P. Kong, C. Ang, Y. Ge, and S. Wen, ``TRITON: high-speed maritime wireless mesh network,'' \emph{IEEE Wireless Commun.}, vol. 20, no. 5, pp. 134--142, Oct. 2013.
\bibitem{BLUECOM}
R. Campos, T. Oliveira, N. Cruz, A. Matos, and J. M. Almeida, ``BLUECOM+: Cost-effective broadband communications
at remote ocean areas,'' in \emph{Proc. OCEANS 2016}, Shanghai, China, Apr. 2016, pp. 1--6.
\bibitem{Inmarsat_5}
P. J. Hadinger, ``Inmarsat Global Xpress: The design, implementation, and activation of a global Ka-band network,''
in \emph{Proc. AIAA Intern. Commun. Sat. Sys. Conf. \& Exhibition}, Queensland, Australia, Sep. 2015,
pp. 1--8.
\bibitem{Iridium_NEXT}
Iridium next satellite constellation. [Online]. Available: http://multivu.prnewswire.com/mnr/iridium/44300/docs/
44300-Iridium\_NEXT\_Brochure.pdf.
\bibitem{Loon}
M. Engel, ``Google's Project Loon hovers over the satellite industry,'' \emph{VIA Satellite}, vol. 28, no. 8, p. 13, Aug.
2013.
\bibitem{zeng_magazine}
Y. Zeng, R. Zhang, and T. J. Lim, ``Wireless communications with unmanned aerial vehicles: opportunities and
challenges,''\emph{ IEEE Commun. Mag.}, vol. 54, no. 5, pp. 36--42, May 2016.
\bibitem{feng_iot}
W. Feng, J. Wang, Y. Chen, X. Wang, N. Ge, and J. Lu, ``UAV-aided MIMO communications for 5G internet of
things,''\emph{ IEEE Internet Things J.}, vol. 6, no. 2, pp. 1731--1740, Apr. 2019.
\bibitem{Cognitive}
E. Lagunas, S. K. Sharma, S. Maleki, S. Chatzinotas, and B. Ottersten, ``Resource allocation for cognitive satellite
communications with incumbent terrestrial networks,'' \emph{IEEE Trans. Cogn. Commun. Netw.}, vol. 1, no. 3, pp. 305--317, Sep. 2015.
\bibitem{wei_lane}
T. Wei, W. Feng, J. Wang, N. Ge, and J. Lu, ``Exploiting the shipping lane information for energy-efficient
maritime communications,'' \emph{IEEE Trans. Veh. Tech.}, vol. 68, no. 7, pp. 7204--7208, Jul. 2019.
\bibitem{air_ground_channel}
D. W. Matolak and R. Sun, ``Air-ground channel characterization for unmanned aircraft systems-Part I: Methods,
measurements, and models for over-water settings,'' \emph{IEEE Trans. Veh. Tech.}, vol. 66, no. 1, pp. 26--44, Jan. 2017.
\bibitem{channel_survey}
C.-X. Wang, J. Bian, J. Sun, W. Zhang, and M. Zhang, ``A survey of 5G channel measurements and models,'' \emph{IEEE
Commun. Surveys Tuts.}, vol. 20, no. 4, pp. 3142--3168, Fourthquarter 2018.
\bibitem{mine_wocc}
X. Li, W. Feng, Y. Chen, C.-X. Wang, and N. Ge, ``Maritime coverage enhancement using uavs coordinated with hybrid
satellite-terrestrial networks,'' \emph{IEEE Trans. Commun.}, vol. 68, no. 4, pp. 2355--2369, Apr. 2020.
\bibitem{zeng_throughput}
Y. Zeng, R. Zhang, and T. J. Lim, ``Throughput maximization for UAV-enabled mobile relaying systems,''
\emph{IEEE Trans. Commun.}, vol. 64, no. 12, pp. 4983--4996, Dec. 2016.
\end{thebibliography}
\end{document}